\documentclass{article}

\usepackage{arxiv}

\usepackage[utf8]{inputenc} 
\usepackage[T1]{fontenc}    
\usepackage{hyperref}       
\usepackage{url}            
\usepackage{booktabs}       
\usepackage{amsfonts}       
\usepackage{nicefrac}       
\usepackage{microtype}      
\usepackage{lipsum}
\usepackage{graphicx}
\usepackage{natbib}
\usepackage{amsmath}
\usepackage{hyperref}
\usepackage{xcolor}
\usepackage{subfig}

\graphicspath{ {./images/} }

\title{Room dimensions and absorption inference from room transfer function via machine learning}

\author{
 Yuanxin Xia \\
  Acoustic Technology, DTU Electro\\
  Technical university of Denmark \\
  Ørsteds Pl. 352, Kongens Lyngby 2800, Denmark\\
  \texttt{xiayuanxin98@gmail.com} \\
   \And
 Cheol-Ho Jeong \\
  Acoustic Technology, DTU Electro\\
  Technical university of Denmark \\
  Ørsteds Pl. 352, Kongens Lyngby 2800, Denmark\\
  \texttt{chje@dtu.dk} \\
}

\begin{document}
\maketitle
\begin{abstract}
The inference of the absorption configuration of an existing room solely using acoustic signals can be challenging. This research presents two methods for estimating the room dimensions and frequency-dependent absorption coefficients using room transfer functions. The first method, a knowledge-based approach, calculates the room dimensions through damped resonant frequencies of the room. The second method, a machine learning approach, employs multi-task convolutional neural networks for inferring the room dimensions and frequency-dependent absorption coefficients of each surface. The study shows that accurate wave-based simulation data can be used to train neural networks for real-world measurements and demonstrates a potential for this algorithm to be used to estimate the boundary input data for room acoustic simulations. The proposed methods can be a valuable tool for room acoustic simulations during acoustic renovation or intervention projects, as they enable to infer the room geometry and absorption conditions with reasonably small data requirements.
\end{abstract}

\section{Introduction \label{ch1}} 
Estimating a room's dimensions and sound absorption configuration without visual cues is a challenging task for human beings owing to the limited capability of the human auditory sensory system. As sound travels very fast, the reflections off the surfaces in three-dimensional spaces overlap even within a short time from sound generation, and thus the human auditory system cannot discriminate each reflection well. In fact, a recent study has shown that listeners often struggle to accurately determine their position in a room based on acoustics alone, even when there are clearly audible differences between positions \citep{meyer2022clearly}. Identifying the sound absorption distribution is not considered important on a daily basis, but it gets more important in practical room acoustic engineering works. For example, in renovations of room acoustics, acousticians need to know the current absorption configuration to diagnose which surfaces cause major problems in the current setting. Throughout such projects, acousticians would need to simulate several absorption configurations to find optimum locations and amounts of absorbers and scattering objects. Therefore, an inference of the absorption configuration from a limited number of measured acoustic signals, typically impulse responses, will be quite useful.

There are several methods proposed to inversely estimate the room geometry from room impulse responses \citep{dokmanic2011can}. One basic idea is to use lower-order strong reflections to estimate the times and directions of arrival. Utilizing multiple source-receiver pairs via various algorithms makes it possible to estimate the geometry of a room \citep{tervo20123d,park2021iterative}. Generally, these analyses are conducted in the time domain using broad-band impulse responses or reflectograms from geometrical acoustics simulations, e.g., the image source method, where waves are simplified as a bundle of straight rays. Therefore wave phenomena, such as diffraction and interference, are often neglected in simulated reflecetograms.

The rapid advancement of machine learning techniques in various engineering fields has significantly enhanced the pattern recognition capabilities of machines. This has inspired researchers to explore the potential of utilizing these approaches to identify patterns within acoustic signals too. \citet{perez2019machine} presented a machine learning-based method for estimating reverberation time from room geometry based on a dataset of measurements. \citet{yu2021room} developed neural networks for inverse estimation of the room size, using time domain room impulse responses (RIRs) as input and the room geometry as output without pre-processing. Their data were generated using the energy-based image-source method, generating approximately half a million RIRs. It must be noted that such energy-based methods can generate reflectograms that include the simplest information of the times of arrival (TOA) and the magnitude of energy at each TOA. These data are easy to use when training a model, but it has several serious drawbacks of not containing phase information and wave phenomena. Therefore, such models trained with energy-based reflectogram dataset are likely to exhibit a limited performance in real-life scenarios, due mainly to lack of wave phenomena and phase information. 

\citet{foy2021mean} conducted a more thorough investigation of estimating the area-weighted absorption coefficient. They also utilized energy-based geometrical acoustics methods combining the image source method and diffuse-rain method. In terms of machine learning, they created neural networks to use the time domain signal as input without a feature extraction step. The output was the absorption coefficient in octave bands, with the error being defined as the absolute difference between the target and estimated values.

Time-domain RIRs as well as reflectograms from the geometrical acoustics simulations, are particularly useful in rooms with nearly rigid walls, as it is relatively straightforward to implement peak detection algorithms and utilize mathematical models to estimate the room geometry or dimension. However, this approach becomes less useful when the room departs from the assumption of rigid walls. In rooms, the number of reflections increases as a cubic function of time, so a heavy overlap among early reflections is observed at times as early as 10 ms after a direct sound according to the mixing and temporal overlap theory \citep{jeong2010room}. 
For non-rigid surfaces, the reflected waves are most likely phase-sifted, leading to a change in the acoustical volume of the room. It is obvious that the higher the absorption in the room, the more imprecise the estimation of the geometry or dimensions, when RIRs or reflectograms are used.

In contrast, a transfer function (TF) in the frequency domain, the Laplace transform pair of a corresponding RIR, includes more precise information about the amount of absorption via two different ways: a frequency shift from the theoretical eigenfrequency assuming a rigid boundary and the half-power bandwidths of the peaks. For a room with perfectly rigid walls, the room dimensions can be obtained by directly applying the relation between the room dimensions and the eigenfrequencies. For a room with sound-absorbing materials, there are methods to restore the room dimensions and deduce the sound absorption configuration.  

The primary focus of this paper is to estimate the low to mid-frequency absorption configuration and extract the room dimensions from room transfer functions. We begin by analyzing the characteristics of damped room modes and peak detection in transfer functions. We employ a supervised machine learning approach to extract the absorption configuration of the rooms.

This paper is structured as follows: Section \ref{ch2} provides an overview of transfer functions in enclosures, including the Helmholtz equation, damped natural frequencies, boundary conditions, the finite element method, and loudspeaker response, with a particular emphasis on the uncertainties. Section \ref{ch3} introduces conventional signal processing and machine learning methods for the room acoustic inference intended. The process of dataset generation and the training strategy are described in Section \ref{ch4}. In Section \ref{ch5} and \ref{ch6}, results and conclusions are presented.

\section{Theory \label{ch2}} 
\subsection{Transfer function in enclosures}
A sound wave propagating within a room, can be represented mathematically using the Helmholtz equation when a source is introduced on the right-hand side as below:

\begin{equation}
	\nabla \hat{p}(\boldsymbol{r}, \omega)+k^2 \hat{p}(\boldsymbol{r}, \omega)=-j \omega \rho Q(\boldsymbol{r}, \omega),
	\label{eq: helmholtz equation}
\end{equation}

where $\hat{p}$ is the sound pressure, $\boldsymbol{r}$ is the position of the source point, $\omega$ is the angular frequency, $\rho$ is the air density, $k$ is the wave number, and $Q(\boldsymbol{r}, \omega)$ describes the source term. For a given source and receiver position, a transfer function is the solution to the Helmholtz equation for the source-receiver pair. The poles of the TF are the eigenfrequencies of the room. In particular, for a rectangular room with dimensions $L_x$, $L_y$, and $L_z$ with rigid boundaries, the eigenfrequencies can be expressed as shown in \autoref{eq: room resonance}, where $n_x$, $n_y$, and $n_z$ are integers ranging from zero to infinity, and $c$ is the speed of sound.

\begin{equation}
	f_{n_x,n_x,n_z}=\frac{c}{2}\sqrt{(\frac{n_x}{L_x})^2+(\frac{n_y}{L_y})^2+(\frac{n_z}{L_z})^2}.
	\label{eq: room resonance}
\end{equation}
\subsection{Damped natural frequencies \label{ch2.2}}
The transfer function of a multi-degree-of-freedom system (MDOF), which is the case in room acoustics, can be represented as a superposition of the responses of single-degree-of-freedom systems (SDOF) \citep{tohyama2011sound,tohyama2015waveform}. The natural frequency of a damped system, $\omega_M$, is expressed as:

\begin{equation}
\omega_M=\sqrt{\omega_d^2-\delta_0^2}=\sqrt{\omega_0^2-2 \delta_0^2}
\end{equation}

where $\omega_0$ is the eigenfrequency, and $\omega_d$ is the angular frequency for
the damped free oscillation \cite[Equation 9.11]{tohyama2015waveform}. The damping constant $\delta_0$ for an eigenfrequency of the room relates to a reverberation time $T_R$ as $\delta_0=\frac{3 \ln 10}{T_R}$. The reverberation time is defined as $T_R=6\ln10\tau_m$, where $\tau_m$ is the time constant of $m^{th}$ resonance, its reciprocal $1/\tau_m$ is equal to the half-power bandwidth $\Delta \omega_m$ of the corresponding resonance \citep{jacobsen2013fundamentals}. Therefore, the reverberation time can also be written as $T_R=\frac{6\ln10}{\Delta{\omega_m}}$, combining the equations gives the expression of damping constant as:

\begin{equation}
\delta_0=\frac{3 \ln 10}{T_R} = \frac{\Delta \omega_m}{2}
\label{eq:DampingConstant}
\end{equation}

The above derivation reveals that for the $m_{th}$ resonance, it is possible to calculate its eigenfrequency $\omega_0^m$ from the damped resonance frequency $\omega_M^m$, if the half-power bandwidth $\Delta {w_m}$ can be found. The equation gives below:
 
 \begin{equation}
 	\omega_0^m = \sqrt{(\omega_M^m)^2+\frac{(\Delta {\omega_m})^2}{2}}
 	\label{eq:undampedres}
 \end{equation}
 
Thus the eigenfrequencies of a rectangular room with absorption treatments depend on the amount of reverberation in the room. The eigenfrequencies at low frequencies are well-separated, but overlap with increasing frequency. When the frequency is high enough, as typically indicated by the Schroeder frequency \citep{schroeder1962frequency}, which indicates a three-fold modal overlap.
\subsection{Boundary conditions: surface impedance and absorption}

The acoustic boundary conditions are typically characterized by three quantities: sound absorption coefficient, surface impedance, and pressure reflection coefficient. The sound absorption coefficient is an energy-based quantity representing how much energy is dissipated by the surface. The surface impedance and pressure reflection coefficients are complex-valued, thus containing phase information. The absorption coefficient is commonly used in the geometrical acoustics method, whereas the other two are used in pressure-based methods.
When modeling porous absorbers, various empirical models are available. A comparison study \citet{oliva2013sound} concluded that Allard and Champoux's model \citep{allard1992new} is the most precise, especially at lower frequencies. Therefore, the present study utilizes the Allard-Champoux model to obtain the surface characteristic impedance ($z_m$) and propagation constant ($\gamma$). In this study, we assumed porous materials are directly placed on a hard surface. The locally reacting assumption is utilized, and \autoref{eq: transmission matrix} provides the surface impedance via a transfer matrix method between the porous material surface and the rigid backing, where the impedance of the latter is represented by $z_c$ and $h$ is the thickness of the porous layer.

\begin{equation}
	Z_s=z_m \frac{z_c \cosh (\gamma h)+z_m \sinh (\gamma h)}{z_c \sinh (\gamma h)+z_m \cosh (\gamma h)}.
	\label{eq: transmission matrix}
\end{equation}

The theoretical random incident absorption coefficient averages the angle-dependent absorption coefficient over the elevation angle $\theta$ from 0 to $\pi/2$  \citep{kuttruff2016room}. However, this method relies on the assumption of an infinitely large surface, which is not a good representative of most real-life scenarios, particularly in small rooms. Therefore, a size correction to the absorption coefficient is introduced assuming the local reaction material \citep{jeong2013converting}, as follows: 

\begin{equation}
	\alpha_{\text {size }}=2 \int_0^{\pi / 2} \dfrac{4 \operatorname{Re}\left(Z_s\right)}{\left|Z_s+\bar{Z}_r\right|^2} \cdot \sin (\theta) d \theta
	\label{eq: size correction ab}
\end{equation}
where $Z_s$ is the surface impedance and $\bar{Z_r}$ is the averaged radiation impedance over the azimuthal angle from 0 to $2\pi$. 

The computation of theoretical radiation impedance requires multiple integrations, which can be computationally demanding and challenging to use. To overcome this, \citet{davy2014average} proposed an approximation method that covers the entire frequency range, adopted in the current study.
\subsection{Finite-element method for data generation}
TFs are generated using commercial numerical software COMSOL ${\mathrm{Multiphysics}}^{\circledR}$ with Livelink$^{\mathrm{TM}}$ for MATLAB$^{\circledR}$. The governing equation with the initial condition and boundary condition in the Finite-element method (FEM) is shown below:

\begin{equation}
	\left\{\begin{array}{l}
		\nabla \cdot\left(-\frac{1}{\rho}\left(\nabla p_{\mathrm{t}}-\mathbf{q}_{\mathrm{d}}\right)\right)-\frac{k^2 p_{\mathrm{t}}}{\rho}=Q, \\
		-\mathbf{n} \cdot\left(-\frac{1}{\rho}\left(\nabla p_{\mathrm{t}}-\mathbf{q}_{\mathrm{d}}\right)\right)=-p_{\mathrm{t}} \frac{i \omega}{Z_{{s}}}.
	\end{array}\right.
\end{equation}

where $p_t = 1$ (Pa), and $\mathbf{q}_{\mathrm{d}}$ is the particle velocity. Note that the frequency-dependent surface impedance $Z_s$ in \autoref{eq: transmission matrix} is used as the boundary condition. The second-order (quadratic) Legendre polynomials are employed for discretization. During the meshing process, the default size is set to one-fifth of the wavelength of the highest study frequency.
\subsection{Loudspeaker response}
Simulations are performed under ideal conditions, e.g., an omni-directional and flat-spectral sound source. However, a loudspeaker in most measurements has a specific directivity pattern and non-flat frequency response, resulting in uncertainties, which was studied in \cite{thydal2021experimental}. In order to enable the simulation to be a good representation of the real-world measurement, it is essential to incorporate available spectral and directivity information. Particularly in lower frequencies, the mismatch between measured and simulated TF is primarily attributed to the loudspeaker's frequency response, which typically exhibits a band-pass characteristic. Therefore, to augment the data generated by the numerical simulation, the loudspeaker's frequency response is compensated for in this study.
\section{Methods \label{ch3}} 
To infer the dimensions and absorption configuration, two  methods are attempted. The first method is mainly based on room acoustic knowledge on the damped eigenfrequencies to estimate the dimensions of the room. The reason why the absorption distribution is not estimated is that the theory in Sec. \ref{ch2.2} relies solely on the averaged reverberation decay, from which estimating the individual absorption characteristics of the surfaces is mathematically undetermined when the measurement points are scarcer than the number of surfaces.

The second method is a black-box machine learning algorithm that can estimate both the dimensions and absorption configuration by training a model with many simulated TFs and data augmentation using the frequency response of the loudspeaker used in real-life.
\subsection{Conventional knowledge-based  signal processing \label{ch3.1}}
This method consists of two steps, eigenfrequency restoration from TFs, and inference of the room dimensions.
 \subsubsection{Eigenfrequency restoration}
In measurements and simulations, it is not uncommon for the frequency resolution to be too sparse, resulting in a bias and difficulty in precisely determining the peak positions of the transfer function, i.e., eigenfrequencies. Numerous methods have been developed to enhance the resolution of frequency response, with Gaussian interpolation being considered one of the most precise methods \citep{gasior2004improving}.  Once the peaks are restored via the Gaussian interpolation, a peak detection algorithm, such as the "findpeaks" command in MATLAB$^{\circledR}$, is used to determine the eigenfrequencies.

\begin{figure}[h]
 	\centering
 	\includegraphics[width=.5\linewidth]{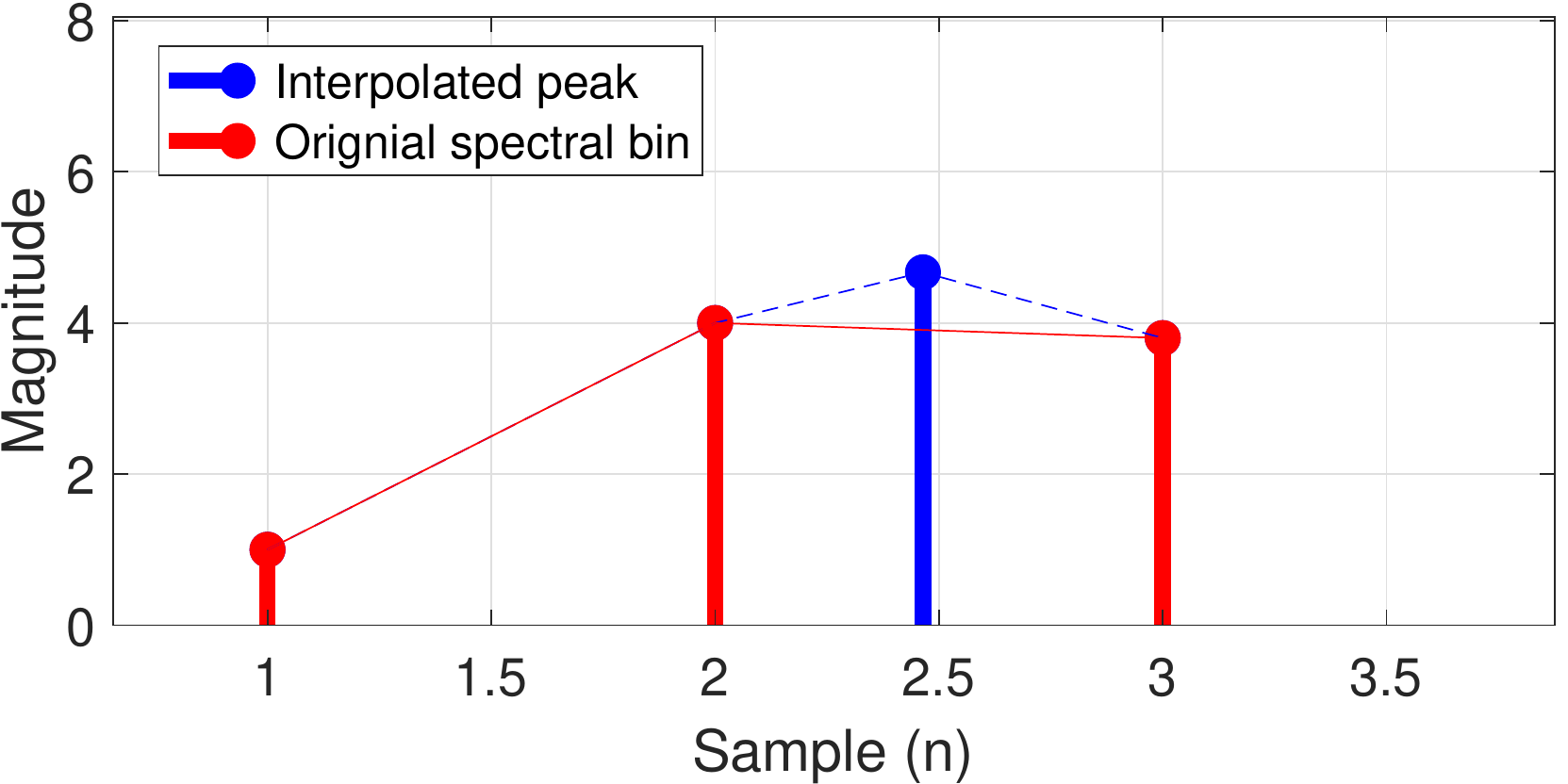}
 	\caption{Gaussian peak interpolation example: discrete frequency points $x_1,x_2,x_3 = 1,2,3$ and corresponding spectral bin $y_1,y_2,y_3 = 1,4,3.8$. The spectral peak maximum is found: $x_{max} = 2.46$ and $y_{max} = 4.67$.}
 	\label{fig:peakinterpolation}
 \end{figure}

Gaussian interpolation is a technique used to estimate the position of a spectral peak maximum situated between two discrete spectral bins by fitting a Gaussian shape. In practice, this method necessitates three discrete frequency points in close proximity to the target, as illustrated in \autoref{fig:peakinterpolation}. 

 \subsubsection{Room dimension inference}
The peaks in the restored room transfer function represent the damped resonance frequency $\omega_M$. The axial mode is described by \autoref{eq:axialmode} and depends on the room dimensions.  
 \begin{equation}
 	\begin{aligned}
 		f(n_x > 0,n_y=0,n_z = 0)&=\frac{c}{2}\frac{n_x}{L_x}\\
 		f(n_x = 0,n_y> 0,n_z = 0)&=\frac{c}{2}\frac{n_y}{L_y}\\
 		f(n_x = 0,n_y=0,n_z > 0)&=\frac{c}{2}\frac{n_z}{L_z}
 	\end{aligned}
 	\label{eq:axialmode}
 \end{equation}
 
 The room dimension can be calculated through the damped resonance frequency $\omega_M$ and half-power bandwidth $\Delta \omega$. For instance, the first-order axial mode (damped resonance) $f_M^1$ and its corresponding half-power bandwidth $\Delta f^1_m$ (in Hz) can be used to determine the dimension of the corresponding axis by the following equation:

 \begin{equation}
 	L = \frac{c}{2\sqrt{(f_M^1)^2+\frac{(\Delta f_m^1)^2}{2}}}.
 	\label{eq:roomsize}
 \end{equation}
 
The challenge here is that we do not have prior knowledge of which peaks correspond to the axial modes. The strategy in this study is to check if there are harmonics and integer multiples of the assumed fundamental axis mode. It is highly unlikely that tangential and oblique modes, 2D and 3D modes, respectively, do have multiples as eigenfrequencies.

 \subsubsection{Limits of conventional methods}
The quality factor (Q-factor) or $\Delta f_m$ measures the peakedness of an individual resonance, which is related to the acoustic damping at the corresponding frequency. Regression analysis can be used to potentially determine the absorption coefficient in a room by measuring the Q-factor of a mode. Again it is cumbersome to determine the absorption coefficient of several surfaces using averaged sound decay information.

\subsection{Supervised learning in machine learning}
The second method (machine learning) takes the TFs as input and the room dimensions and individual absorption coefficients as output by regressing a function to learn patterns hidden in the transfer functions using neural networks. This section uses multi-task convolutional neural networks to be designed and optimized based on room acoustic knowledge. 

\subsubsection{Principles}
The aim of neural networks is to extract information from TFs. Specifically, convolutional neural networks (CNNs) are considered for feature extraction with TFs as the input and the inference of room dimension (length, width, and height) and absorption coefficients of each surface for the frequency bands centered at 63 Hz, 125 Hz, and 250 Hz, as output, which is often described as multi-task learning (MTL).

\begin{figure}[h]
	\centering
    \includegraphics[width = .5\linewidth]{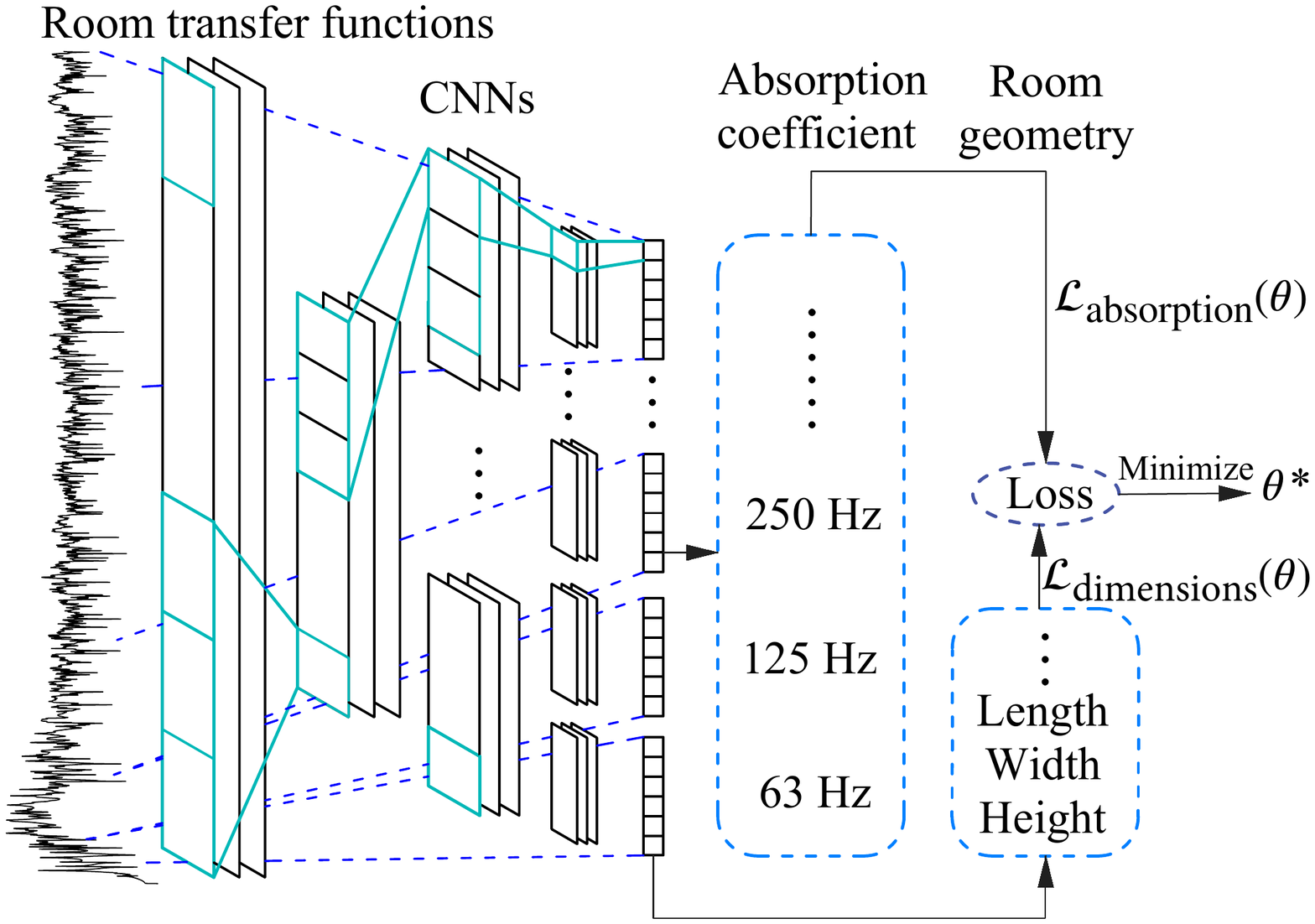}
	\caption{Implementation of CNNs constrained by room dimensions and absorption. The room transfer function is the direct input for the feature extraction process. Each part of the input signals corresponds to a frequency band, and the room dimensions are integrated into the lowest frequency band. The loss function of the network consists of two components: the room dimensions and the absorption coefficients.}
	\label{fig: Neural network concept}
\end{figure}

The design of multi-task neural networks aims to optimize information sharing across multiple tasks. The methodology for determining task branches is informed by the calculation method for the single value of the absorption coefficient. By dividing the task into several frequency bands, each branch can be specifically allocated to features directly related to the absorption coefficient of the corresponding frequency band, as illustrated in \autoref{fig: Neural network concept}. At each frequency band, not only one absorption coefficient be estimated, but it could also give the absorption coefficient of each surface in the targeting room if we assign multiple tasks.

In multi-task learning, variations in noise patterns among tasks can be mitigated by incorporating inductive bias. This approach allows the multiple tasks to act as regularizers for one another, thereby reducing the risk of overfitting. As a result, the model is able to achieve improved representation by averaging the noise patterns. As mentioned, the Q-factor is related to the absorption coefficient. Thus, identifying the position of each resonance may help neural networks understand the absorption distribution. Although the room size is associated with all the resonances of the room transfer function, the axial mode is exclusively associated with $L_x$, $L_y$, and $L_z$ as shown in (\autoref{eq:axialmode}). The lowest three eigenfrequencies are axial modes unless the room is not highly disproportionate, so the branch of room size is integrated into the lowest octave band branch. This allows sharing room dimension information with the lowest octave band 63 Hz, in particular.

\subsubsection{Loss function formulation}
To enable the deep learning structure to follow the proposed constraints, we consider a set of TFs as $\boldsymbol{\mathrm{x}}$, corresponding to labelled absorption coefficients $\boldsymbol{\mathrm{y}}_a$ and room dimension $\boldsymbol{\mathrm{y}}_d$. Although the TFs in a single room  vary with the source and receiver position, the proposed CNNs are able to detect the room information regardless of the variation of positions to some extent.

 The loss function is composed of two components, namely the room dimension and absorption coefficient loss, $\mathcal{L}_{\text {dimension}}$ and $\mathcal{L}_{\text {absorption}}$. The total loss is technically a weighted sum of the dimension loss and absorption loss at each octave band as follows:

 \begin{equation}
\mathcal{L}(\mathrm{\theta})=\mathcal{L}_{\text {dimension}}(\mathrm{\theta})+\mathcal{L}_{\text {absorption}}(\mathrm{\theta}).
\end{equation}

where

\begin{equation}
    \mathcal{L}_{\text {dimension}}(\mathrm{\theta}) = \lambda_d\sum_{i=1}^{N_d}(y_d^{(i)}-f_{\boldsymbol{\theta}_d}^{(i)}(\boldsymbol{\mathrm{x}}))^2,
\end{equation}

\begin{equation}
    \mathcal{L}_{\text {absorption}}(\mathrm{\theta})=\sum_{j=1}^{N_f \cdot N_s}\lambda_a^{(j)}(y_a^{(j)}-f_{\boldsymbol{\theta}_a}^{(j)}(\boldsymbol{\mathrm{x}}))^2.
\end{equation}

Here, $\lambda_d$ denotes a single value weight of room dimension. If we consider a shoe-box-shaped room, $N_d$ is three, corresponding to the width, length, and height. For the absorption coefficient loss, $\boldsymbol{\lambda}_a$ is a vector consisting of the weights for the octave band absorption coefficients. $N_f$ is the number of frequency bands and $N_s$ is the number of surfaces in the room.

\section{Dataset generation and hyperparameter tuning \label{ch4}} 
\subsection{Dataset}
The  absorption coefficients in room acoustic engineering are typically necessary from 125 Hz to 4 kHz according to ISO 3382-1 \citep{ISO3382-1}. However, computations up to such high frequencies using wave-based numerical methods are expensive. The conventional inference method described in Sec \ref{ch3.1} does not hold when the modal overlap is high. Therefore, we limited the frequency range for the TF dataset to 250 Hz octave band. The size-corrected random-incident absorption coefficients of each surface over three-octave bands, 63 Hz, 125 Hz, and 250 Hz are calculated, and the length, width, and height are labeled as a label series for one room transfer function.

\subsubsection{Material properties}
We will use measurements from \citep{thydal2021experimental} to infer the room dimension and absorption properties, as it describes all the details about the room and materials sufficiently well. The paper used  two types of porous materials: A being Ecophon Akusto Wall-A (40 mm glass wool with a specific flow resistivity of 47,000 $\mathrm{Ns/m^4}$) and B being Ecophon Industry Modus (100 mm glass wool with a specific flow resistivity of 109,000 $\mathrm{Ns/m^4}$). The absorption coefficient of the remaining concrete surfaces is set to [0.029, 0.048, 0.043] for the 63 Hz, 125 Hz, and 250 Hz octave bands, respectively.

\subsubsection{Variation of material configuration and room dimensions}
\begin{figure}[h]
	\centering
	\includegraphics[height = .2\linewidth]{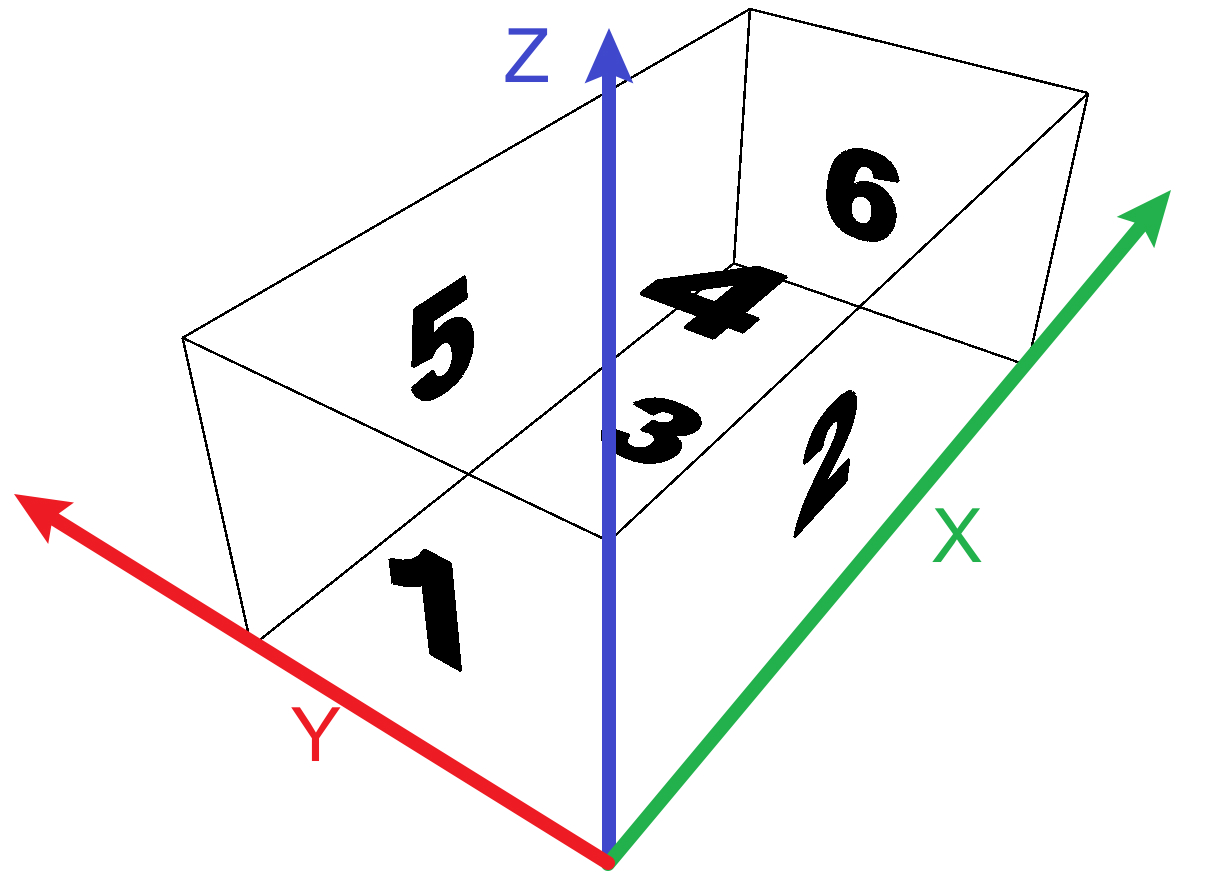}
	\caption{Surface number assignment.}
	\label{fig: Surface number}
\end{figure}

In \autoref{fig: Surface number},  a rectangular room is shown with numerical labels assigned to the surfaces. To generate the training data, material configurations are varied according to \autoref{tab: Material arrangement}, with a dash indicating the concrete surface.

\begin{table*}[h]
	\centering
	\begin{tabular}{c|c|c|c|c|c|c|c|c|c|c|c|c|c}
		\toprule
		\text { Surface } & 1 & 2 & 3 & 4 & 5 & 6 & \text { Surface } & 1 & 2 & 3 & 4 & 5 & 6 \\
		\hline \text { Config 1 } & \textbf{ A } & - & - & - & - & - & \text { Config 15 } & - & \textbf{A} & - & - & \textbf{A} & - \\
		\text { Config 2 } & \textbf{ B } & - & - & - & - & - & \text { Config 16 } & - & \textbf{B} & - & - & \textbf{B} & - \\
		\text { Config 3 } & - & \textbf{A} & - & - & - & - & \text { Config 17 } & \textbf{A} & - & - & - & - & \textbf{A} \\
		\text { Config 4 } & - & \textbf{B} & - & - & - & - & \text { Config 18 } & \textbf{B} & - & - & - & - & \textbf{B} \\
		\text { Config 5 } & - & - & \textbf{A} & - & - & - & \text { Config 19 } & \textbf{A} & - & - & \textbf{A} & - & \textbf{A} \\
		\text { Config 6 } & - & - & \textbf{B} & - & - & - & \text { Config 20 } & \textbf{B} & - & - & \textbf{B} & - & \textbf{B} \\
		\text { Config 7 } & - & - & - & \textbf{A} & - & - & \text { Config 21 } & \textbf{A} & - & - & \textbf{B} & - & \textbf{A} \\
		\text { Config 8 } & - & - & - & \textbf{B} & - & - & \text { Config 22 } & \textbf{B} & - & - & \textbf{A} & - & \textbf{B} \\
		\text { Config 9 } & - & - & - & - & \textbf{A} & - & \text { Config 23 } & - & \textbf{A} & \textbf{A} & - & \textbf{A} & - \\
		\text { Config 10 } & - & - & - & - & \textbf{B} & - & \text { Config 24 } & - & \textbf{B} & \textbf{B} & - & \textbf{B} & - \\
		\text { Config 11 } & - & - & - & - & - & \textbf{A} & \text { Config 25 } & - & \textbf{A} & \textbf{B} & - & \textbf{A} & - \\
		\text { Config 12 } & - & - & - & - & - & \textbf{B} & \text { Config 26 } & - & \textbf{B} & \textbf{A} & - & \textbf{B} & - \\
		\text { Config 13 } & - & - & \textbf{A} & \textbf{A} & - & - & \text {Rigid }&- & -&- & -&-&- \\
		\text { Config 14 } & - & - & \textbf{B} & \textbf{B} & - & - &\text { Summary } & 8 & 8 & 8 & 8 & 8 & 8 \\
		\bottomrule
	\end{tabular}
	\caption{Material assignment of each surface.}
	\label{tab: Material arrangement}
\end{table*}

\begin{table*}[h]
\centering
\begin{tabular}{cccccc}
\toprule
{ Room Ratio}               & { Length(m)} & { Width (m)} & { Height (m)} & { Area ($\mathrm{m^2}$)} & { Volume ($\mathrm{m^3}$)} \\ \hline
{ 1:1.11:1.67}              & { 3.0}       & { 4.5}      & { 2.7}       & { 13.50}    & { 36.45}       \\\
{ $\sim$Bolt (2:3:5)}       & { 4.0}       & { 6.75}     & { 2.7}       & { 27.00}    & { 72.90}       \\
{ $\sim$Louden (1:1.4:1.9)} & { 3.8}       & { 5.15}     & { 2.7}       & { 19.47}    & { 52.84}       \\
{ $\sim$Cox(1:1.56:1.86)}   & { 4.2}       & { 5.0}      & { 2.7}       & { 21.00}    & { 56.70}       \\
{ 1:1.33:2.66}              & { 4.0}       & { 8.0}      & { 3.0}       & { 32.00}    & { 96.00}       \\
{1:1.67:1.67}               & { 5.0}       & { 5.0}      & { 3.0}       & { 25.00}    & { 75.00}       \\
{1:1:1.3}                   & { 4.3}       & { 3.3}      & { 3.3}       & { 14.19}    & { 46.83}       \\
\bottomrule
\end{tabular}
\caption{Aspect ratio of the room.}
\label{tab: room dimension arrangement}
\end{table*}

As shown in \autoref{tab: room dimension arrangement}, the training dataset primarily consists of small rooms, and the aspect ratios are determined based on the research of \citet{appolloni2021housing}, which outlines general room aspect ratios across Europe. Research in optimizing rectangular room dimensions has been conducted to distribute modal frequencies uniformly and prevent degenerate modes, which are also considered in designing the room dimensions \citep{bolt1946note,louden1971dimension,cox2004room}.
\subsubsection{Variation of source and receiver position}

\begin{figure}[h]
	\centering
	\subfloat[]{\includegraphics[width=0.25\linewidth]{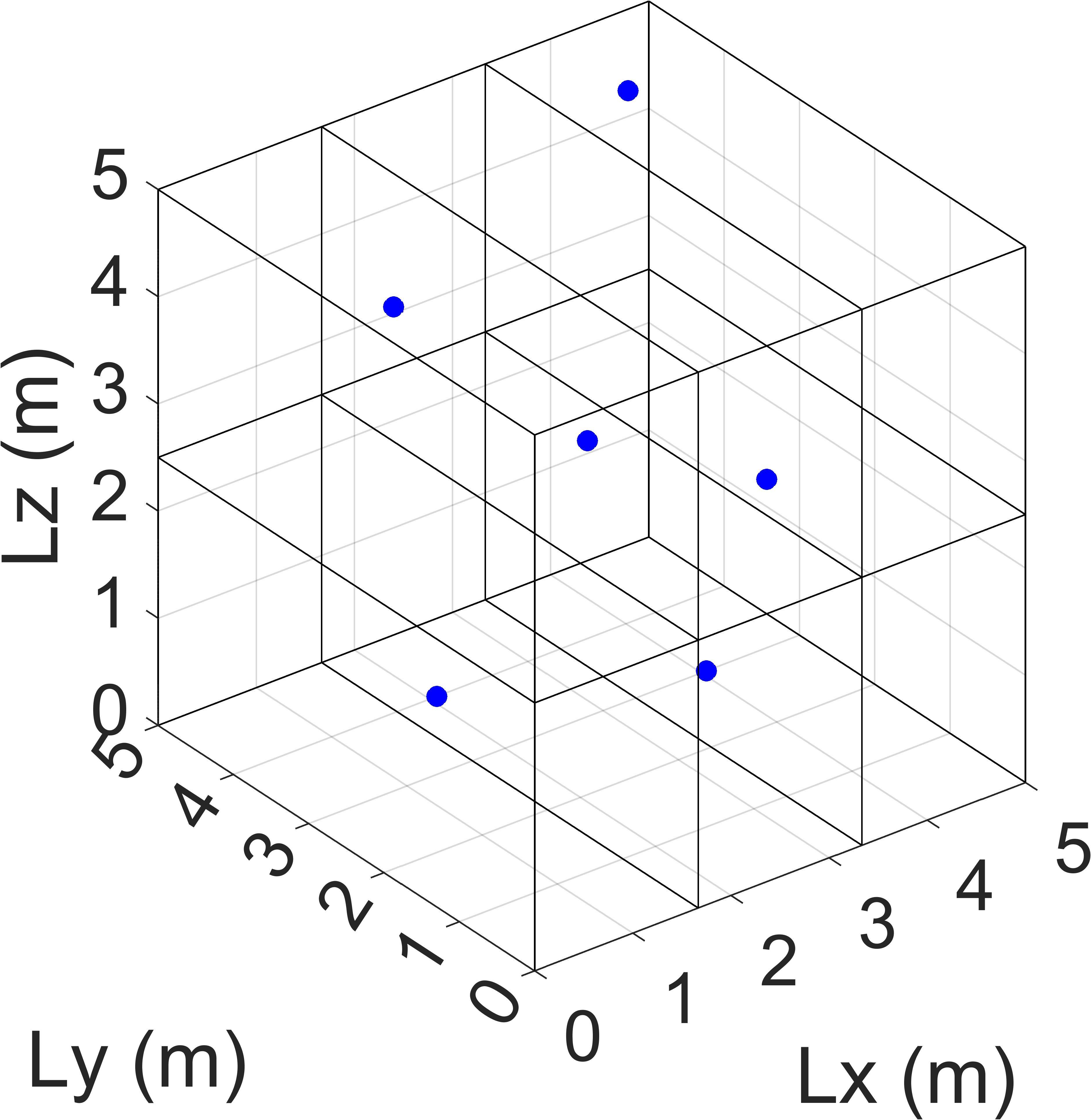}}
	\subfloat[]{\includegraphics[width=0.25\linewidth]{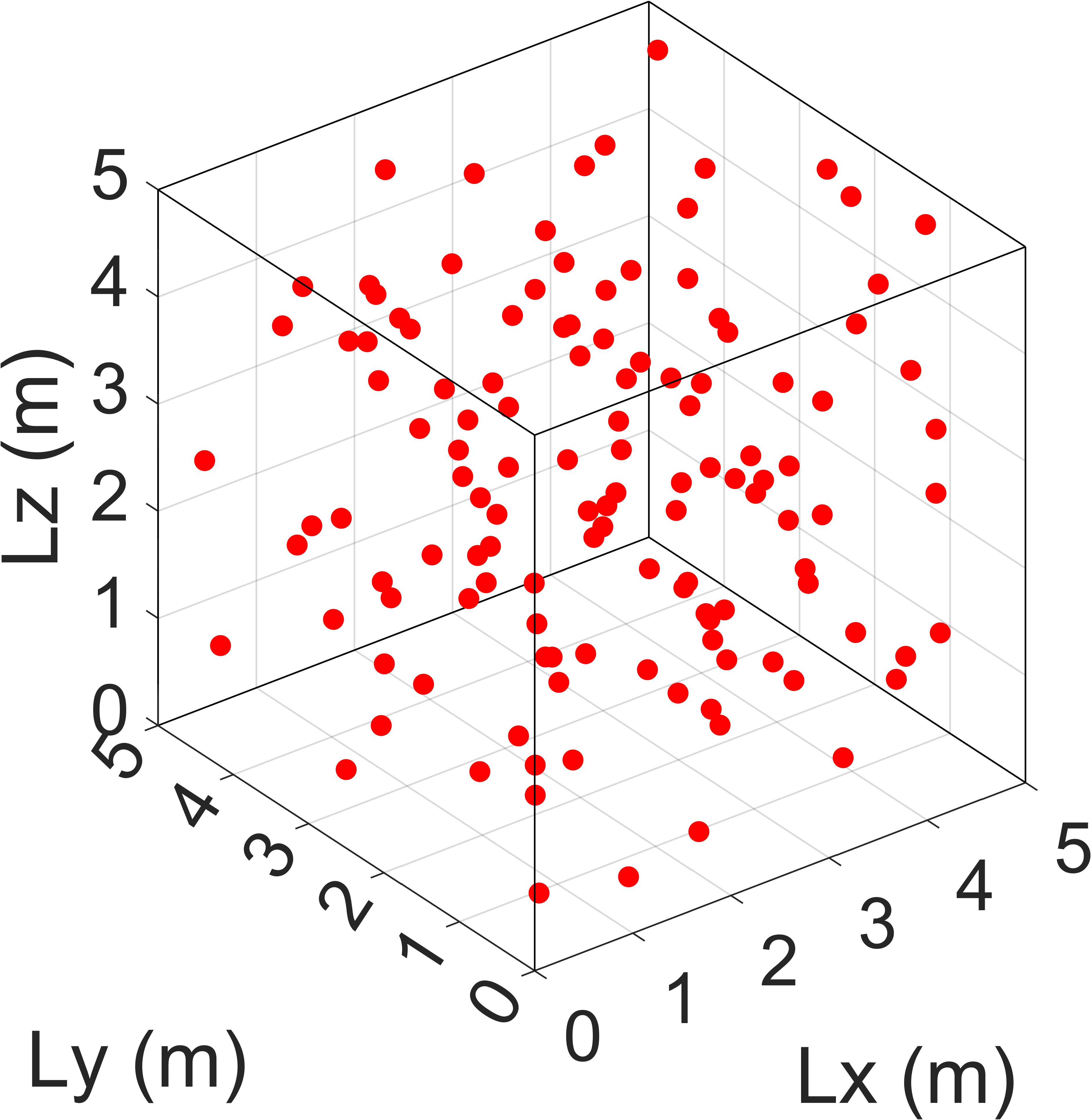}}
	\caption{Example of the distribution of sources (a) and receivers (b) in a room.}
	\label{fig:sourcedis}
\end{figure}

 The Latin hypercube sampling is used to optimize the positioning of sound sources and receivers in a room to improve computational efficiency and ensure a well-distributed sampling of the room's transfer function. To prevent the influence of nodal lines on room mode capture, as depicted in \autoref{fig:sourcedis}(a), the rectangular room is divided into two layers, each further divided into three subspaces, resulting in six smaller subspaces for random distribution of individual sources. For receiver points, the room is divided into 125 (5x5x5) smaller spaces using the same method as shown in \autoref{fig:sourcedis}(b), with receivers randomly distributed within each subspace. Randomly chosen sound sources and receivers are used to calculate in each room to generate the TF dataset.

\subsubsection{Source spectral correction}

The sources are assumed to have a flat frequency response in the simulation. However, real-world loudspeakers can be modeled as band-pass filters, allowing a limited range of frequencies to pass. They exhibit resonances at low frequencies, boosting the radiation at certain frequencies and reducing low-frequency responses. To ensure the feasibility of our model for real-world measurements, the simulated dataset is augmented through a correction based on the on-axis spectra of various dodecahedra sources measured in an anechoic using \citet{moos2021loudspeaker}. 

\subsubsection{Dataset summary}
In all simulations, the sound speed is fixed at 343 $\mathrm{m/s}$ and the air density is 1.20 $\mathrm{kg/m^3}$. The dataset comprises 7 rooms of varying dimensions in \autoref{tab: room dimension arrangement} , each with 27 material configurations in \autoref{tab: Material arrangement}, 6 different sources, and 125 evenly spaced receiver positions. Thus, the dataset includes a total of $7\times 27 \times 6 \times 125 = 141,750$ transfer functions. The frequency resolution of the transfer functions is 0.5 Hz, and the frequency ranges from 1 to 354 Hz. The simulation dataset is abbreviated as "Simu.", while the source corrected dataset is abbreviated as "Aug." below.
\subsection{Neural networks structure}
The standard ResNet V2 architecture, as described in \citep{he2016identity}, is utilized as the backbone model due to its superior performance for optimizing loss function \citep{li2018visualizing}. ResNet V2 possesses multiple variants with varying depths, the selection of which is contingent upon the complexity of the task at hand. A deeper model possesses a greater ability for nonlinear expression and can learn more intricate transformations, thus accommodating more complex feature inputs. However, it should be noted that a deeper network also poses greater optimization challenges. According to \citet{nichani2020increasing}, an increase in the neural networks depth can decrease the test error, but the error increases beyond a certain threshold. Thus, it is crucial to match the depth of the networks to the complexity of the problem at hand. Through experimentation with various depths of ResNet architecture, it was found that ResNet18 achieved the best performance, regardless of the filter widths.

The activation function for the absorption coefficient outputs in each layer is chosen as the sigmoid function, providing results in the range of 0 to 1, aligning with the definition of the absorption coefficient. The activation function for the room dimension output layer is the rectified linear unit (ReLU), resulting in outputs between 0 and infinity.

\begin{figure}[h]
	\centering
    \includegraphics[width = .5\linewidth]{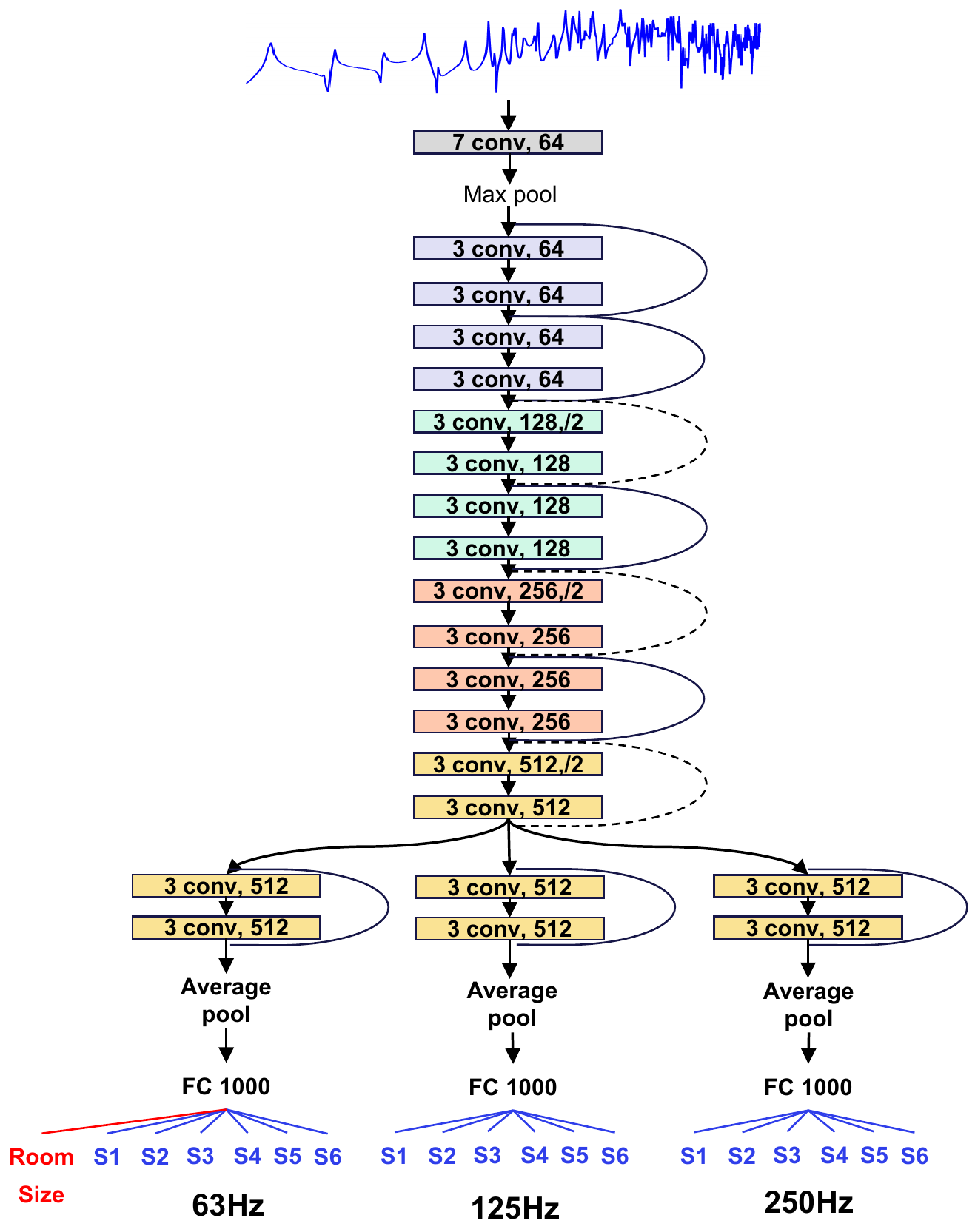}
	\caption{Multi-task residue networks for absorption coefficient and room dimension. In the output layer, the blue line represents the sigmoid activation function, and the red line represents ReLu activation function.}
	\label{fig: Multi-task residue networks}
\end{figure}

\subsection{Training parameter}
In this study, a multi-task neural network was built using the TensorFlow framework (\autoref{fig: Multi-task residue networks}). Hyperparameter selection utilized popular choices found in recent deep learning literature \citep{he2019bag}. The training parameters, including a 0.8:0.2 split for training and testing data, respectively, a batch size of 32 and 200 epochs, were employed. Adam optimization \citep{kingma2014adam} algorithm was utilized with an initial learning rate of $1\times 10^{-4}$. The number of filters in the input layer was set to 64, and the remaining network structure adhered to the general principles of a residual network. Batch normalization \citep{ioffe2015batch} was applied post-convolution to prevent overfitting and He initialization, as described in \citep{he2015delving}, was used for weight initialization. All experiments were executed on the NVIDIA Tesla V100 GPU.

\subsubsection{Loss weights of absorption coefficient}

The method for determining loss weights in MTL is still an open area of research. Various methods have been proposed, yet there is no established understanding or universal design for loss weighting in MTL. A commonly used strategy is the uniform distribution of losses across tasks, which serves as a baseline for many proposed methods. Other approaches include the uncertainty weighting method, which assigns weights based on the noise pattern of individual tasks \citep{kendall2018multi}, and the dynamic weight average method, which employs an attention mechanism to take into account task difficulty and prevent the domination of easier tasks \citep{Liu2019CVPR}. It is important to note that the overall loss is dominated by the small gradient term in MTL, hence more difficult tasks should receive a larger share of the loss weight to balance the overall loss. In the current study, the room dimension was considered as an easier task compared to the absorption coefficient estimation task, which is a highly abstract feature of the room transfer function, and thus more difficult to predict. Furthermore, the difficulty of predicting the absorption coefficient was found to increase with frequency. As a first experiment, a uniform loss weight was assigned to all tasks, and the room dimension branch was removed to eliminate any potential disturbance.

\begin{figure}[h]
	\centering
	\includegraphics[width=.5\linewidth]{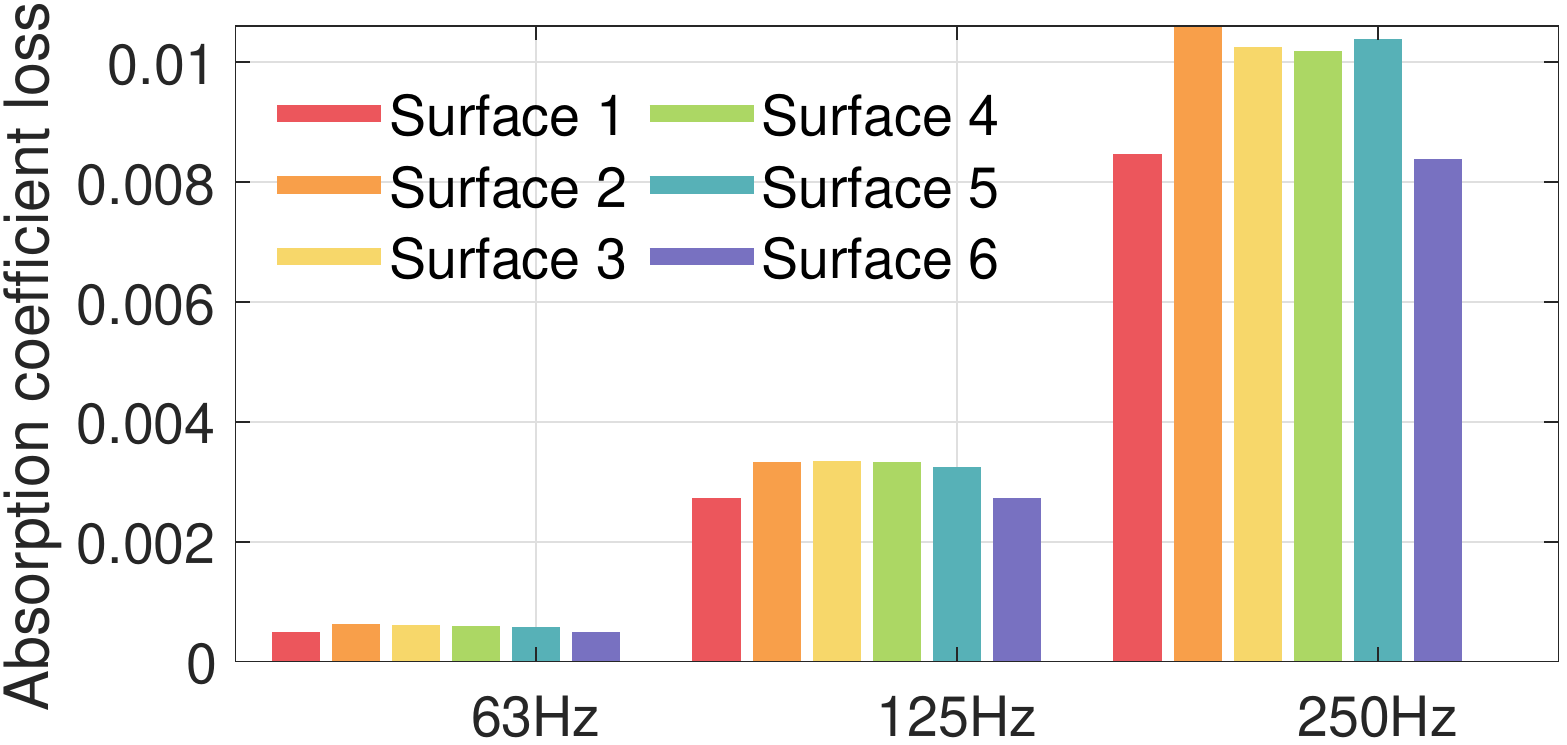}
	\caption{Estimation loss of the absorption coefficient as a function of frequency for the testing dataset. The results show that the loss increases with frequency, indicating that feature extraction becomes more challenging at higher frequencies.}
	\label{fig:absorptioncoefficientloss}
\end{figure}

As depicted in \autoref{fig:absorptioncoefficientloss}, the loss of the absorption coefficient increases linearly with the frequency bandwidth, in accordance with the constant percentage bandwidth rule. The goal of the neural networks is to provide precise predictions for all absorption coefficients, regardless of the frequency band. Therefore, it is important to balance the loss of the room absorption coefficient by prioritizing difficult tasks. One approach is to assign weights to the absorption coefficient that are proportional to the frequency bandwidth. Specifically, the weight of the neighboring higher-frequency octave band is twice as high as that of the current octave band.

In previous experiments, the weights for the absorption losses were equally assigned a value of 1, resulting in a total weight of 18. To maintain consistency, the sum of all weights for the absorption losses will remain as 18 in the following experiments. This weighting strategy, referred to as "Frequency-Based Weights" (FBW), involves assigning higher weights to the absorption loss for higher frequencies. For example, the weight for the absorption coefficient at 63 Hz is calculated as $\lambda_{63}=3/7$, while the weights for the absorption coefficients at 125 Hz and 250 Hz are $6/7$ and $12/7$, respectively.

\subsubsection{Loss weight of the room dimension}
The weight for the dimension loss in the MTL structure depends on the task and the difficulty of estimating it compared to the absorption coefficients. The weight for the dimension loss can be adjusted during the training process to improve its prediction accuracy. Our approach is to keep the sum of the absorption loss weights as 18 and compare the weight for the dimension loss to this sum, the results of this approach can be seen in Experiments 7 to 12 in \autoref{fig:TrainingStrategy}.

\subsubsection{Frequency resolution and training accuracy}
The frequency resolution of a TF can be a crucial factor in the training of neural networks. While interpolation and other techniques can improve the frequency resolution of the original data, we can improve the frequency resolution during the TF simulations. However, a finer resolution increases the computational resources further. We assess the impact of frequency resolution by varying it to 0.5 Hz in the original simulation and 1 Hz by down-sampling. The results are compared and summarized in \autoref{fig:TrainingStrategy}.

\subsubsection{Experiment design and result}
\begin{figure*}
    \centering
    \includegraphics[width=1\linewidth]{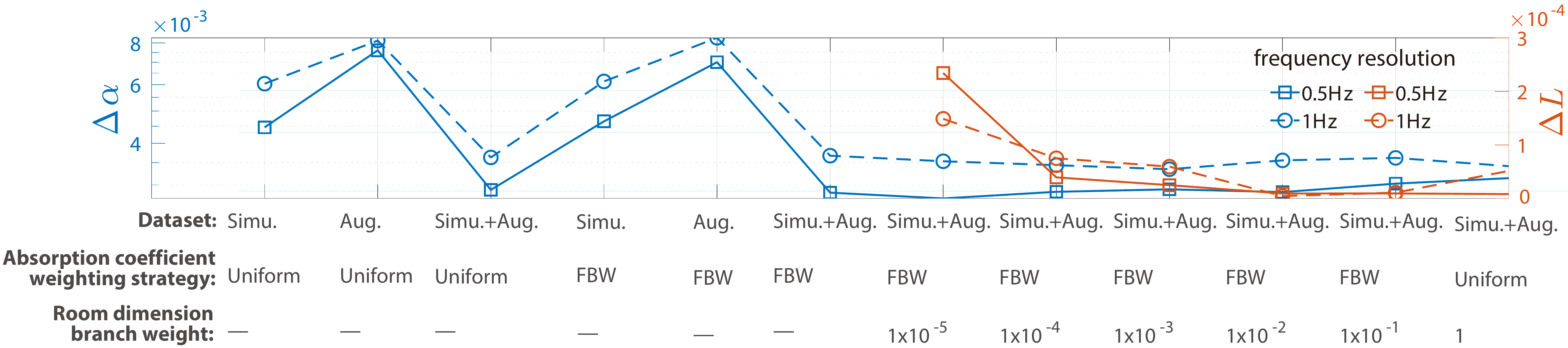}
    \caption{Comparison of different training strategies. The blue line represents the naive sum of the absorption coefficient inference error across surfaces and frequencies on the testing dataset. The combination of simulation and augmented data can significantly improve the model performance, regardless of the weighting strategy. The red line represents the naive sum of the room dimension inference error. A higher weight on the room dimension leads to a lower inference error.}
    \label{fig:TrainingStrategy}
\end{figure*}

Twelve experiments were conducted to evaluate the proposed training strategy. 
Twelve cases consist of different dataset being {Simu}, {Aug}, and {Simu+Aug}: flat-response source vs. spectral correction to the sound source, the absorption loss weighting to vary from {uniform} to {FBW}: uniform weights vs. frequency-based weights, and the dimension loss weight adjusted to [$10^{-4}$,$10^{-3}$,$10^{-2}$,$10^{-1}$,$1$] as described in the $x$-label of \autoref{fig:TrainingStrategy}.

To evaluate the performance of the multi-task neural network and quantify the errors in absorption coefficient and room size estimations, two error metrics were introduced. These error metrics allow us to compare the effectiveness of different training strategies and configurations. The first error metric, denoted as $\Delta \alpha$, measures the average difference between the estimated and true absorption coefficients across all surfaces and frequency bands, as defined by the equation:

\begin{equation}
\Delta \alpha = \frac{1}{N_f\cdot N_s}\sum\limits_{i=1}^{N_s}\sum\limits_{j=1}^{N_f}(\alpha_{i,j}^{est}-\alpha_{i,j}^{true})^2
\end{equation}
where $N_f$ is the number of frequency bands, $N_s$ is the number of surfaces, and $\alpha_{i,j}^{est}$ and $\alpha_{i,j}^{true}$ represent the estimated and true absorption coefficients, respectively.

The second error metric, $\Delta L$, evaluates the average difference between the estimated and true room dimensions across all dimensions (length, width, and height), defined by:

\begin{equation}
\Delta L = \frac{1}{N_d}\sum\limits_{i=1}^{N_d}(L_i^{est}-L_i^{true})^2
\end{equation}
where $N_d$ represents the number of dimensions and $L_i^{est}$ and $L_i^{true}$ correspond to the estimated and true room dimensions, respectively.

As expected, the frequency resolution of 0.5 Hz outperforms the 1 Hz resolution. Between Simu, Aug, or Simu+Aug, the Sim+Aug significantly outperforms the other two cases. Between uniform absorption loss weights and FBW, the absorption error $\Delta \alpha $ is smaller with the FBW than with the uniform weights.

Between the two frequency resolutions compared, the 1 Hz resolution dataset degraded the accuracy compared to the 0.5 Hz resolution case, as expected. However, the overall decrease in the performance of the absorption coefficient prediction was within 0.01, which is considered to be small. A potential error introduced by improper dataset augmentation, i.e., spectral source correction, was greater than the error by the sparse frequency resolution. These results suggest that the frequency resolution of the 1 Hz is sufficient to counterbalance the computational load in generating the training data, and a correct augmentation is more important for improving the overall performance of the neural networks.

\section{Validation using measurements \label{ch5}} 
\begin{figure}[h] 
	\centering
        \subfloat[]{\includegraphics[width=0.15\linewidth]{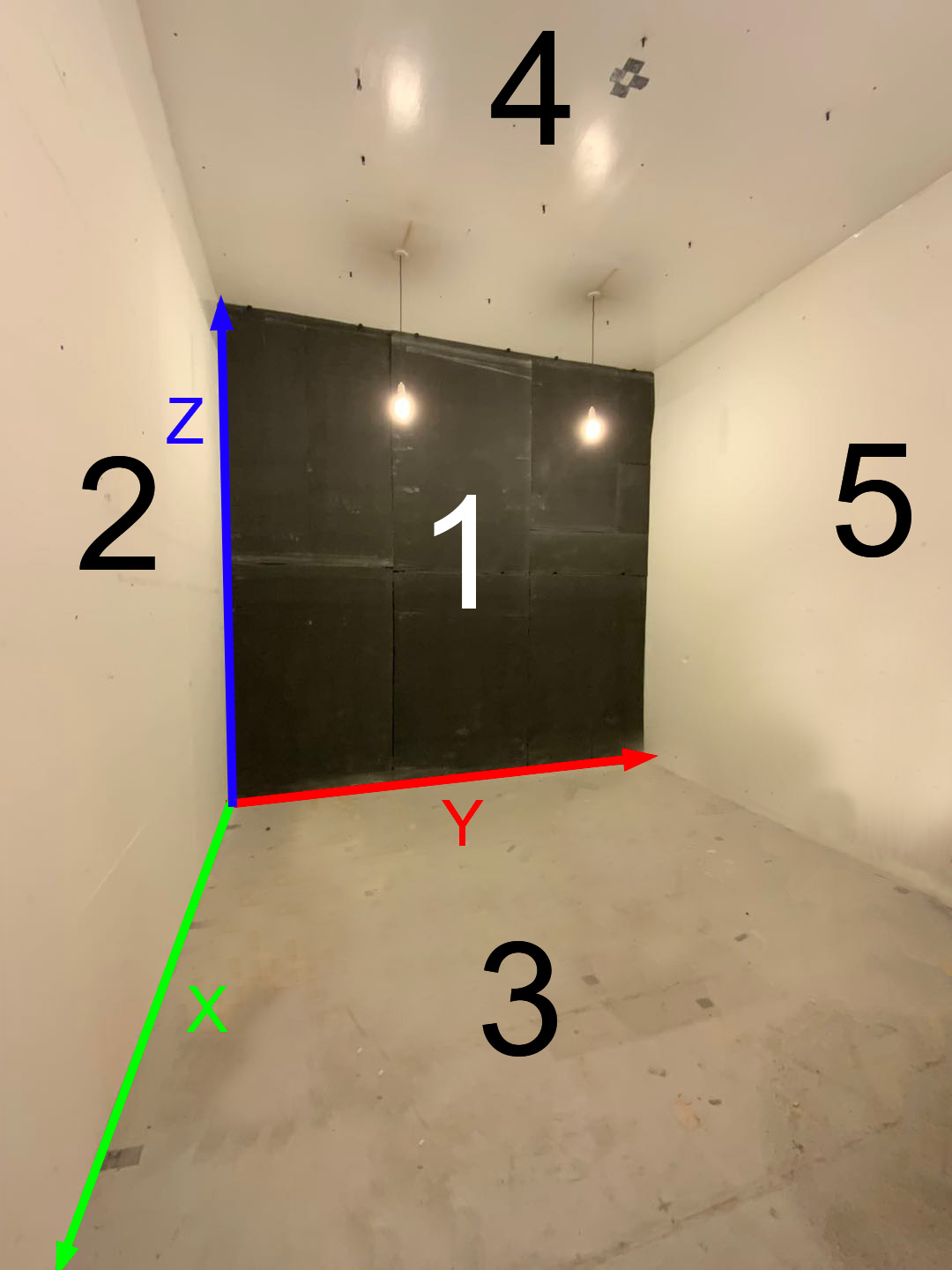}\includegraphics[width=0.15\linewidth]{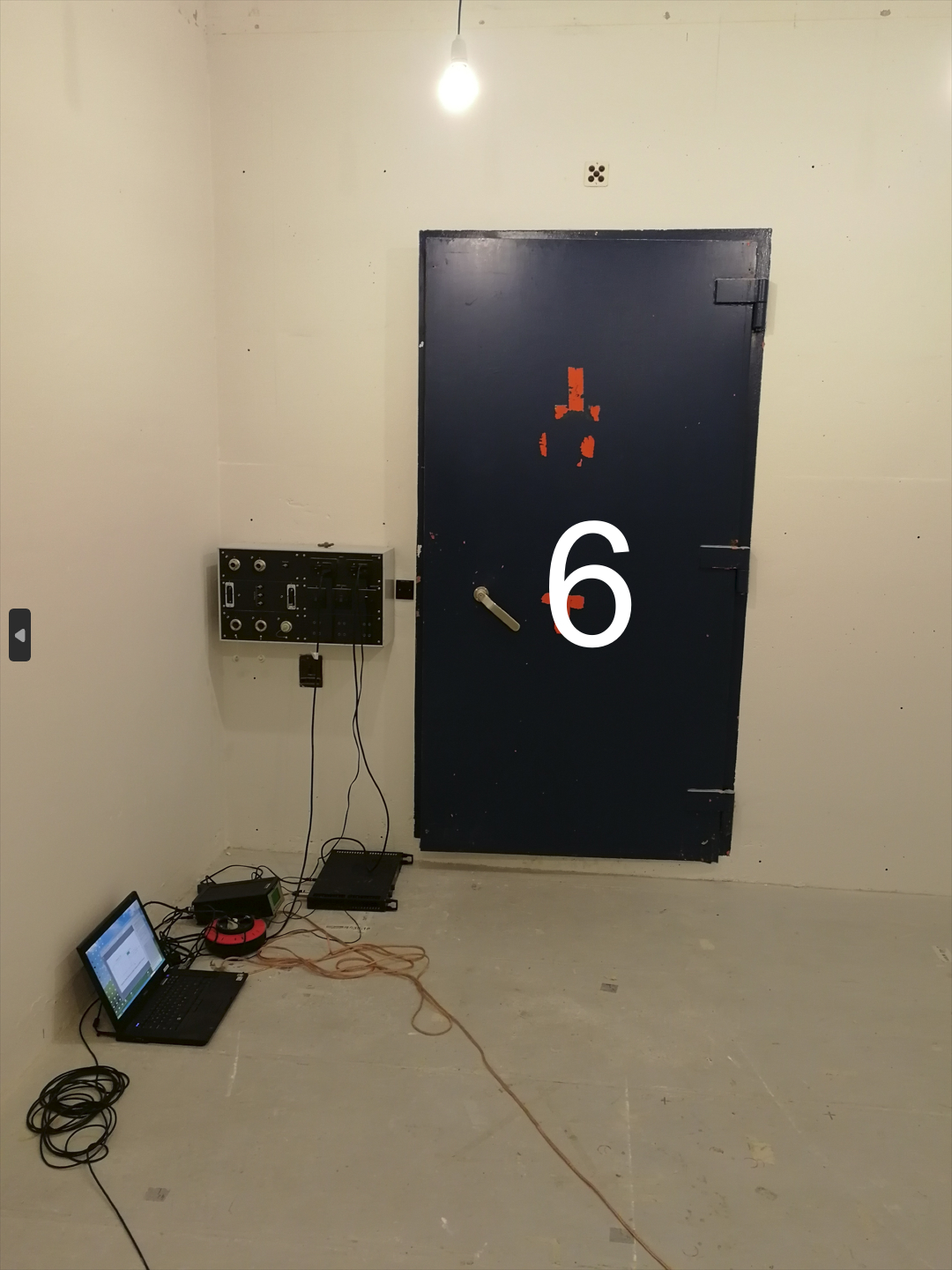}}
        \subfloat[]{\includegraphics[width=0.3\linewidth]{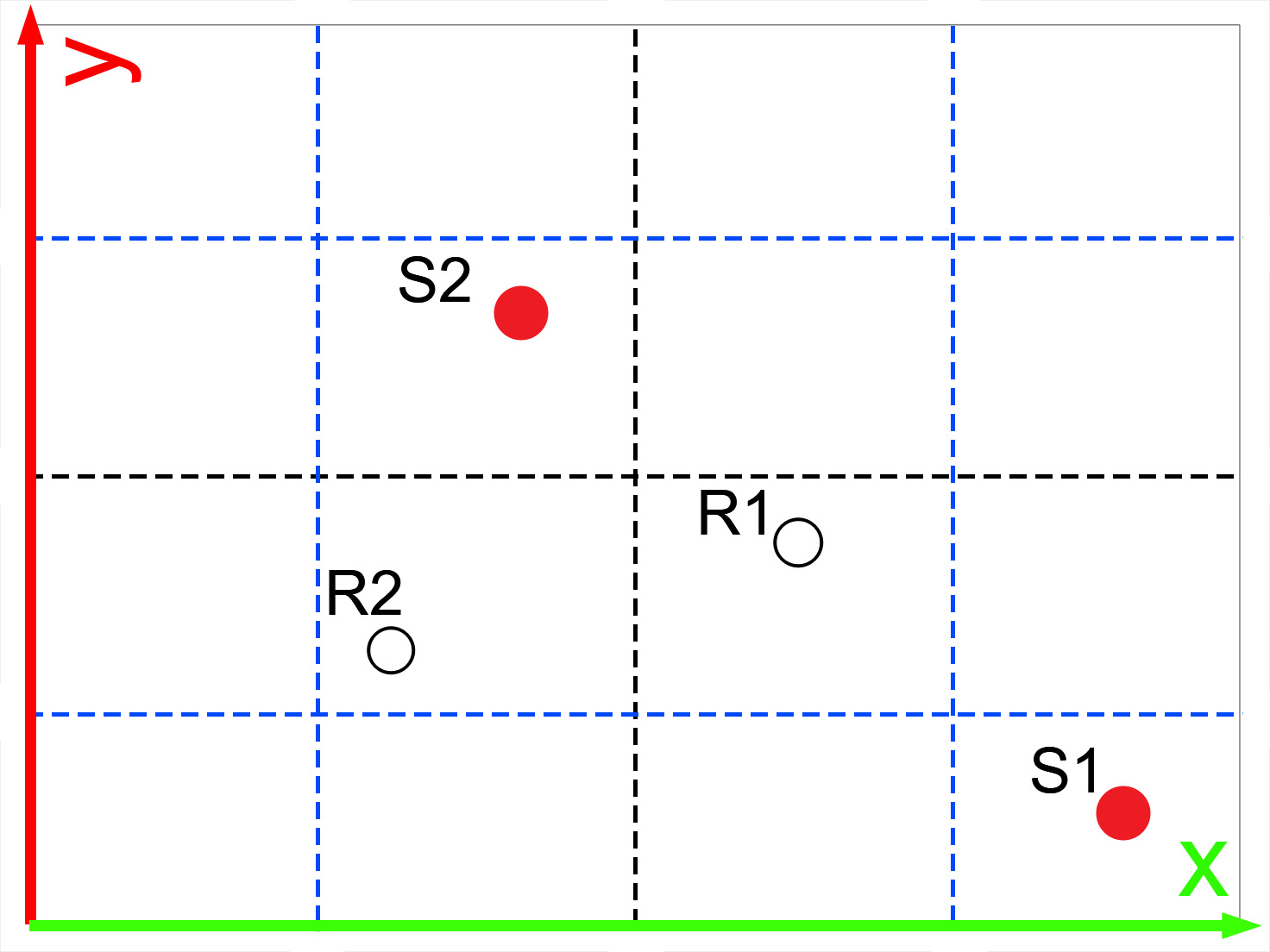}}
	\caption{Measurement setup based on the work in \cite{thydal2021experimental}. The figure includes (a) the coordinate system in the measurement room and (b) the source and receiver positions, as well as the nodal lines. The black dashed line represents the first-order mode, and the blue dashed line represents the second-order mode.}
    \label{fig:testingroom}
\end{figure}

The validation of the proposed model was performed using the measured dataset from \cite{thydal2021experimental}. It is important to note that during the measurement, no acoustic material was present on five sides of the room, and only two types of acoustic materials (Material A and Material B) were installed on surface 1, as illustrated in \autoref{fig:testingroom}(a). During the measurement, both the source and receivers were fixed. The room with source and receivers is shown in \autoref{fig:testingroom}(b). It can be observed that neither the source nor the receiver points are located on the nodal line of the first- and second-order room modes in the $x$ and $y$-directions, which eliminates any concerns about the modes not captured by this measurement. However, Receiver 1 is closer to the nodal line of the first axial mode in the $y$-direction, and this proximity may impact the inference of the room dimensions in the $y$-direction. In the $z$-direction, the height of the source and receiver are 1.6 $\mathrm{m}$ and 1.45 $\mathrm{m}$, respectively, while the room height is 3.29 $\mathrm{m}$. Note that the sound source, an in-house omnidirectional loudspeaker, has a diameter of 0.4 $\mathrm{m}$, not an ideal point source, so the acoustic center may be closer to the nodal line of the first axial mode in the $z$-direction, leading to a reduction in the amplitude of the corresponding mode in the transfer function. 

All measured impulse responses were truncated to 2 seconds, with the Fast Fourier Transform applied to obtain the respective TFs, resulting in a 0.5 Hz frequency resolution. To account for the sound source's pressure level in the simulated data, a normalization process was implemented. As the sound source for simulated data had a pressure level of 1 Pa, the sound pressure level (SPL) was generally 91 dB at 1 Hz. This normalization process aligned the amplitude of the measured and simulated data, with all TFs normalized to 91 dB SPL at 1 Hz. This adjustment enabled a more precise evaluation of the network's performance with real-world measurements.

A series of experiments, outlined in \autoref{fig:TrainingStrategy}, aimed to determine the most effective parameter combination for multi-task neural networks applied to simulation data. Optimal results were achieved using Sim+Aug data, FBW for absorption loss, and a $\lambda_d$ value of $1\times10^{-5}$ on the testing data. This configuration was subsequently chosen for the multi-task neural networks. To enhance the networks' generalization performance with measurement data, the entire simulated dataset was used for training.

The performance of the networks was then assessed using real-world measured TFs as test data. To enable the neural networks to better adapt to the target measured data and address the challenge of domain shift, a domain adaptation process was applied to reduce discrepancies between the training data (simulated) and the new domain (measured). This process entailed fine-tuning the networks by feeding them the measured TF for the S1R1 pair, along with the corresponding labels (Material A, Material B, and rigid), and retraining for 1000 epochs using a full batch size. This allowed the networks to assimilate information from these three specific TFs. Subsequently, the generalization performance of the networks was evaluated for the S1R2, S2R1, and S2R2 pairs.

\subsection{Room dimension inference}
\begin{figure}
    \centering
    \includegraphics[width = .6\linewidth]{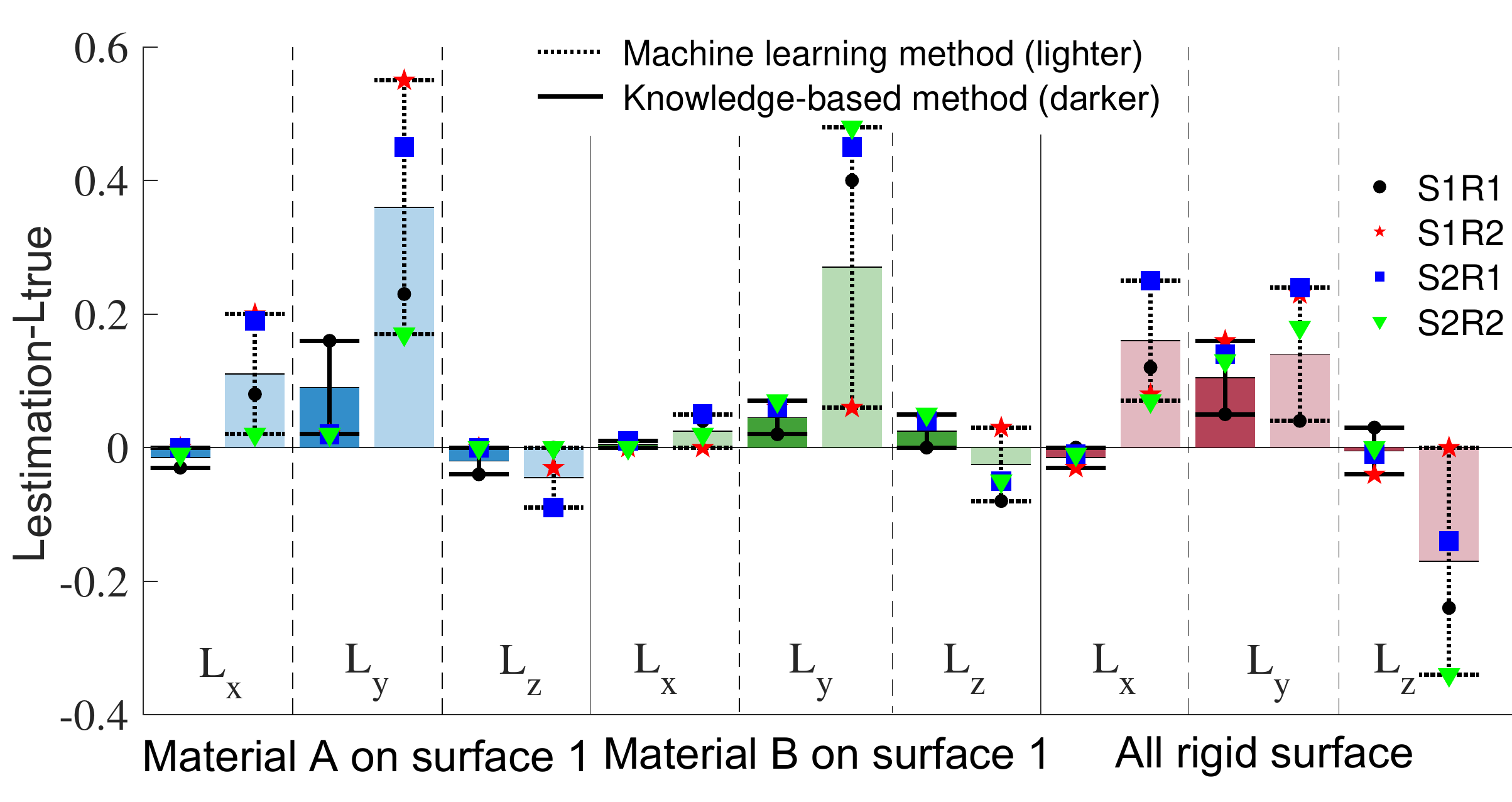}
    \caption{Comparison of dimension inference results using two methods. A positive error means overestimation, while a negative error means underestimation.}
    \label{fig:DimensionError}
\end{figure}

In the knowledge-based conventional method, an algorithm for identifying the axial mode was developed using a function for the restored peak. The function first searches for all peaks in lower frequencies and then look for any corresponding higher-order axial modes (integer multiples of the detected peaks) of these lower-frequency peaks. However, in some cases, this approach does not yield accurate results, necessitating manual selection of the peak to find the exact axial mode. The results of the predictions are shown in \autoref{fig:DimensionError}, which compares the predicted room dimensions to the actual measurements and calculates the errors. The errors were found to be within 10 cm in all directions, which confirms the accuracy of the knowledge-based method, despite the need for manual adjustments in certain instances. It is noted that the $y$-axis direction had the largest error, possibly due to the proximity of the receiver to the corresponding nodal line.

The prediction results by the neural network are also presented in \autoref{fig:DimensionError}. Compared to the knowledge-based approach, the error of the neural network prediction is larger by an order of magnitude. However, the neural network scheme does not require manual parameters optimization, and its robustness is relatively high. The predictions of $L_x$ and $L_z$ are relatively accurate, while the predictions of $L_z$ deviate significantly from the actual values. This trend is observed in the knowledge-based scheme. The $L_z$ is generally underestimated, which may be due to the overestimated $L_y$. Overall, the locations of the sound source and receiver have a greater impact on the neural network performance than the knowledge-based scheme. This is likely due to the feature bias of the measured data and the neural network's vulnerability to extrapolate unseen transfer functions.

\subsection{Absorption coefficient inference}
\begin{figure*}[h]
    \centering
    \includegraphics[width = 1\linewidth]{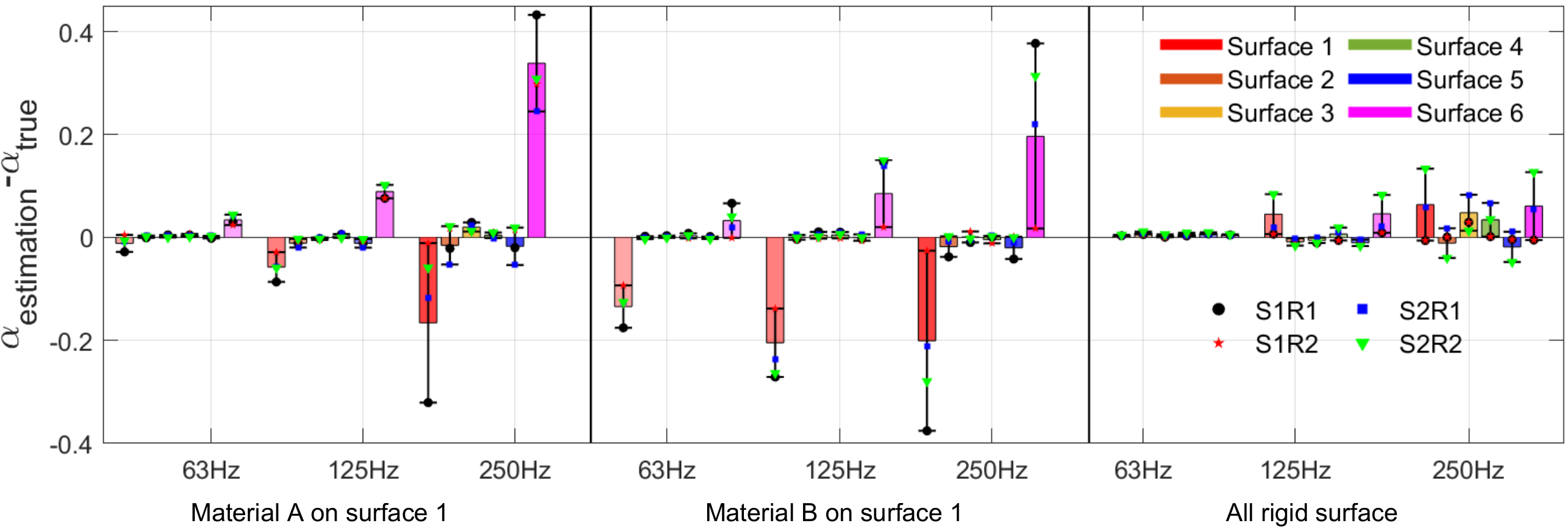}
    \caption{Estimation error of the absorption coefficient for the measured data. The error for the rigid wall configuration increases with frequency. In the absorption material configuration, the model tends to overestimate surface 6 and underestimate surface 1, suggesting that a single RTF input captures the energy flow direction but cannot differentiate between opposite surfaces. The estimation for the remaining surfaces is relatively accurate. The position of source/receiver combinations also affects prediction accuracy, particularly when the pair is far from the absorbed surface (S1R1).}
    \label{fig:AbsorptionError}
\end{figure*}

The dimension and absorption coefficient inference within a limited frequency range up to the 250 Hz octave band, is presented in \autoref{fig:AbsorptionError} via the machine learning scheme. The prediction error increases with frequency in both rigid wall and material configurations. Note that Surface 1 and 6 are parallel to each other. The model tends to overestimate the absorption coefficient for surface 6 and underestimate it for surface 1, suggesting that the model with a single TF input can capture the significantly reduced energy flow along the $x$-direction, but cannot accurately capture exactly which surface has a higher absorption. Nevertheless, the prediction accuracy for the remaining surfaces is acceptable (within the 0.05 round error required by ISO 11654 in the 250Hz octave band), and the position of the source/receiver combinations influences the prediction accuracy, especially when the pair is far from the absorbed surface (S1R1). The relative error of prediction decreases as the material's absorption coefficient increases, and the neural network's prediction result is typically smaller than the theoretical calculation of the absorption coefficient.

\section{Conclusion \label{ch6}} 
This study evaluates the feasibility of two methods to estimate the room dimensions and frequency-independent absorption coefficients, using measured transfer functions up to the 250 Hz octave band. The knowledge-based and supervised machine learning techniques were compared, with a focus on the inference performance of the damped resonant frequency to the room's eigenfrequency in the knowledge-based method and the utilization of a multi-task convolutional neural network in the machine learning approach. The results demonstrate the potential for using simulation data to train neural networks for real-world measurements. The knowledge-based method outperforms the machine learning method in estimating room dimensions. Instead, the machine learning method can estimate the frequency-dependent absorption properties at each surface, whereas the knowledge-based method cannot. The ResNet 18 showed promising results in estimating the dominant directions for the energy decay, but precise estimations of the absorption coefficients of a parallel surface pair were not guaranteed. In this study, the training data generation took 99.66\%, and the model training took 0.34\% of time. Once the model has been trained and loaded, one inference takes around 20ms. The time distribution highlights the computational challenges of generating a large training dataset. Further research should focus on domain adaptation, mitigating the influence of feature shifts between measured and simulated data, and enhancing the generalization performance of neural networks by incorporating physical information and expanding the dataset.

\bibliographystyle{cas-model2-names}  
\bibliography{references}

\end{document}